\documentclass[aps,twocolumn,prl,pre,amssymb]{revtex4}

\usepackage{graphicx}

\begin{document}

\title{\textbf{Model of the electrochemical conversion of an undoped organic
   semiconductor film to a doped conductor film}}

\author{M. Modestov}
\affiliation{
The Plasma and Nonlinear Physics Group,
Department of Physics, Ume{\aa} University, SE--901 87
Ume{\aa}, Sweden}

\author{V. Bychkov}
\affiliation{
The Plasma and Nonlinear Physics Group,
Department of Physics, Ume{\aa} University, SE--901 87
Ume{\aa}, Sweden}

\author{G. Brodin}
\affiliation{
The Plasma and Nonlinear Physics Group,
Department of Physics, Ume{\aa} University, SE--901 87
Ume{\aa}, Sweden}

\author{D. Valiev}
\affiliation{
The Plasma and Nonlinear Physics Group,
Department of Physics, Ume{\aa} University, SE--901 87
Ume{\aa}, Sweden}

\author{M. Marklund}
\email{mattias.marklund@physics.umu.se}
\affiliation{
The Plasma and Nonlinear Physics Group,
Department of Physics, Ume{\aa} University, SE--901 87
Ume{\aa}, Sweden}

\author{P. Matyba and L. Edman}

\affiliation{The Organic Photonics and Electronics Group, Department of Physics, Ume{\aa} University, SE--901 87 Ume{\aa}, Sweden}

\bigskip

\begin{abstract}
We develop a model describing the electrochemical conversion of an organic
semiconductor (specifically, the active material in a light-emitting electrochemical cell) from the undoped non-conducting state to the doped
conducting state. The model takes into account both
strongly concentration-dependent mobility and diffusion for the electronic charge carriers
and the Nernst equation in the doped conducting regions. 
It is demonstrated that the experimentally
observed doping front progression in light-emitting electrochemical cells
can be accurately described with this model. 
\end{abstract}

\maketitle

\newpage
The pioneering demonstration that it is possible to perform doping
on an organic conjugated polymer and attain a high metallic-like
electronic conductivity was awarded with the Nobel Prize in
Chemistry in 2000 \cite{1}. The general opportunity for a
controlled tuning of the electronic and optical properties of
organic semiconductors via various doping techniques has opened up for a wide range of emerging novel
and flexible applications. It also provides for important
and stimulating science as regards to the fundamental processes
in organic semiconductors. In the latter
context, it is relevant to find and understand the
differences and similarities to the well-established physics and
chemistry that is dictating the behavior of conventional inorganic
semiconductors.

A striking difference between organic and inorganic semiconductors
is that the former are soft materials with weak intermolecular bonds
and low dielectric constants, and as such are strongly influenced by
electron-lattice interactions \cite{2}. The mere existence of an excited electronic state on an organic
semiconductor creates a significant local lattice distortion, which
causes the electronic charge carrier to self-localize over a spatial
volume of nm-sized dimensions (and to be termed an electron/hole
polaron). Moreover, local disorder is common  in
 polymers, and the existence of a random distribution of
dopant counter-ions
will create even more energetic disorder and further localize the
electronic charge carriers \cite{3,4}.

A number of interesting features of organic semiconductors originate
in their specific properties and distinguish them from inorganic
semiconductors. First, electrochemical (and also chemical) doping
can be performed with straightforward means\textit{ in-situ}. By
applying an appropriate voltage to an electrode coated with an
organic-semiconductor film and in contact with an electrolyte, it is
possible to inject electronic charge into the organic semiconductor,
which subsequently is electro-statically compensated by injection
(ejection) of ions from (into) the electrolyte \cite{5,6}. The
necessary motion of ions within the organic film is facilitated by
its soft, and in some cases porous, nature. Second, the doping
levels correlating to a high electronic conductivity is much higher
in organic semiconductors than in their inorganic counterparts
($\sim $0.1 vs.\ $\sim $10$^{{\rm -} {\rm 4}}$ dopants/repeat unit),
which is a direct consequence of the self-localization effects in
organic semiconductors and the concomitant positive dependence of
the mobility (\textit{$\mu $}) of electrons and holes on
concentration ($n)$, \cite{4,7,8}. Finally, the mobility \textit{$\mu
$}($n)$ exhibits a strong dependence on the doping mode, and a much
stronger positive dependence is in effect when near-by compensating
(and lattice-polarizing) counter-ions are present, as is the case in
electrochemical and chemical doping, than when the electronic charge
carriers are introduced in, $e.g.$, a field-effect
mode \cite{4,8}.

In this Letter, we develop a model describing the electrochemical conversion
of an organic semiconductor film from the undoped non-conducting state to
the doped conducting state, and compare the results with recently
acquired experimental data. It is demonstrated that the observed doping front
in light-emitting electrochemical cell (LEC) devices can be described with a model based on a combination of a
strongly concentration-dependent mobility for the electronic charge carriers
and the Nernst equation in the doped  conducting
 regions.

 A typical LEC consists of a solid-state active material, comprising
an intimate blend of a fluorescent conjugated polymer and an
electrolyte, positioned between two electrodes. When a voltage equal
to or larger than the band-gap potential of the conjugated polymer
is applied between the two electrodes $(V \geq E_{g}/e)$, balanced
charge injection (electrons at the cathode and holes at the anode)
into the conjugated polymer is facilitated by the preceding
migratory motion of the electrochemically inert ions and the
corresponding formation of electric double layers at the two
electrode interfaces. The injected electronic charge carriers are
subsequently in effect electrostatically neutralized by the
compensatory motion of ions in a process termed electrochemical
doping; p-type doping ($i.e.,$ injection of holes and compensation
by anions) takes place at the positive anode and n-type doping
(injection of electron and compensation by cations) at the negative
cathode, and after a ``turn-on time'' a light-emitting p-n junction
is formed in the inter-electrode gap \cite{9,Nature-Mat}.
The doping of a fluorescent conjugated polymer has a ``dark'' optical
signature in that the formation of dopants (polarons) is concomitant with
the formation of low-energy sites with limited radiative-decay probability
from the excited state ($i.e.$, quenched fluorescence). Accordingly, by exposing
an LEC device during operation in a dark environment to UV light (which
excites the fluorescence of the conjugated polymer), it is possible to
correlate the formation of a doped conjugated polymer to the emergence of
dark regions with quenched fluorescence.
Figure 1a presents a photograph of a planar LEC device during
operation at $V$ = 5 V $(>E_{g}/e)$ and under UV light illumination
in a dark room. The anode/cathode is marked with a ``$+/-$''. The fluorescence quenching of the ``green'' and
the ``red'' components of the photograph along a representative path
in the inter-electrode gap (as indicated by the dotted line in Fig.\
1b) are presented in Figs.\ 1c and d, respectively. Two observed distinct dark
regions with very low fluorescence intensity correspond to p-type
and n-type doping. The regions originate at the
anodic and cathodic interfaces and end with a very sharp and
rather irregularly shaped front boundary.
By correlating the size and growth of each doped region
with the temporal evolution of the integrated current, it is
possible to extract important information as regards to the doping
concentration and profile. We have recently demonstrated that the
doping concentration in the doped regions is very high, with a value
of the order of $\sim $10$^{{\rm 2}{\rm 6}}$ dopants/m$^{{\rm 3}}$
(corresponding to $\sim $0.1 dopants/repeat
unit), \cite{10}, and that the doping concentration behind
the doping front is relatively constant \cite{11,12,13,14}.
%However, it is relevant to point out that the accuracy of this method is
%dependent on that the measured current solely produces
%electrochemical doping. Since we have recently demonstrated that
%electrochemical and chemical side reactions can be rather prominent
%on the cathodic  \cite{10}, \cite{13}
%-- but not the anodic  \cite{14} -- side in this specific system,
%we accordingly,

%%%% FIG 1 %%%%
\begin{figure}
	\includegraphics[width=0.98\columnwidth]{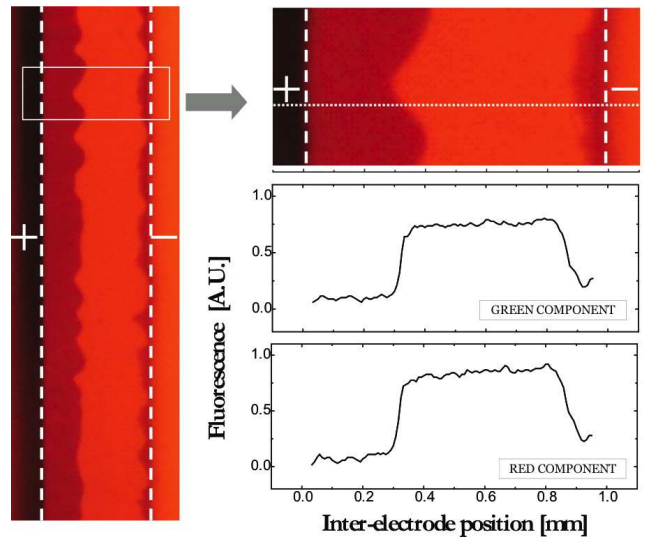}
	\caption{(a) Photograph of a planar Au/\{MEH-PPV+PEO+KCF$_{{\rm 3}}$SO$_{{\rm
3}}$/Au device with a 1 mm inter-electrode gap during operation at $V$ = 5 V
and $T$ = 360 K. The anode and cathode are indicated by a (+) and a (-) sign,
respectively. The device was operated in a dark room under UV illumination,
so that the p-type and n-type doping regions are visualized as dark regions
originating from the anodic and cathodic interfaces, respectively. (b)
Enlarged section of the photograph in (a), as indicated by the solid line.
The dotted line in (b) marks the path in the inter-electrode gap along which
the fluorescence quenching data of (c) the green component and (d) the red
component were extracted from the photograph.}
\end{figure}
%%%%%%%%%%%

The organic semiconductor material will be modeled as consisting of
electrons and holes together with positive and negative ions in a polymer
structure. Diffusion and mobility of charges are of primary importance, which may be studied using the methods of plasma physics.
%An injection barrier will be also introduced in the form of a
%potential, taking into account Fermi level effects and the semiconducting
%nature of the organic material.
Similar models have been presented before, but we will show that very important aspects have
previously been overlooked.
%when deriving the governing dynamical equations for such materials.
In particular, we include an injection barrier in
the form of the Nernst potential in conjunction with a concentration
dependent mobility, which are necessary to obtain agreement between
theory and experiments.
%Moreover, more detailed models including, \textit{e.g.}, kinetic effects, can be introduced but will not change the main results of the present Letter.
The particle species in the material satisfy the force
balance equation
\begin{equation}
\label{eq1}
nm\dot{\bf v} = - qn\nabla (\phi - \phi _{N} ) - k_{B}
T\nabla n - \tau^{-1}nm{\rm {\bf v}},
\end{equation}
where $n$ denotes the concentration, $m$ is the mass,
$\bf{v}$ is the velocity, $q$ is the charge, $k_{B} $ is Boltzmann's
constant, $T$ is the temperature, $\tau $ is the time between
collisions, and $\phi - \phi_{N} $ is the \textit{overpotential}
experienced by the particle. Here we have introduced a
phenomenological injection barrier $\phi_{N} $ due to the difference
in Fermi levels for the electrons and holes between the doped
and  undoped regions.
Ions do not experience any barrier, so that
$\phi_{N}$ should be taken zero for ions. In the electrochemical system under study, the injection process is not limiting \cite{15},
\textit{i.e.}, the electronic species on both sides of the barrier are at
quasi-equilibrium, and $\phi_{N} $ is chosen to be the Nernst potential
\begin{equation}
\label{eq2}
\phi _{N} (n_{h,e} ,T) = (k_{B} T/q_{h,e})\ln \left[ n_{h,e}/(n_{h,e,\infty}  - n_{h,e}) \right],
\end{equation}
where $n_{h,e,\infty}$ is the maximum concentration of holes (label "$h$") or electrons (label "$e$"). The Nernst potential originates from quantum
effects, and the interpretation of $\phi_{N} $ in terms of an electrostatic force should
therefore be done with caution.
%Moreover, the important dynamical aspects of
%the Nernst potential enter Eq.\ (\ref{eq1}) in the form of a gradient.
The Nernst potential is  valid only in the highly
doped region, and it should  be set formally to zero in the
undoped region, $\phi _{N} (n_{h,e}  ,T) = 0$.
When going over from  one region to another,
we match the solutions
by fitting the functions and their derivatives.

On the timescales of interest here, \textit{i.e.}, much longer than the
collision time $\tau $, the collision force dominates over the particle acceleration $ mn \,d {\bf {v}}/ dt$, such that the
left hand side of Eq.\ (\ref{eq1}) can be neglected. Then
${\rm {\bf v}} = - Dn^{-1}\nabla n \pm \mu \nabla (\phi - \phi _{N}
)$,
where the mobility is given by $\mu =\tau \left|q\right|/m$,
%(note the inclusion of the sign of the charge)
 the diffusion coefficient $D$ follows from the
Einstein relation $\mu = \left|q\right|D / k_{B} T$ and $\pm$ correspond to positive and negative charges. Substituting ${\rm {\bf v}}$ into
the continuity equations for all species we obtain the equations for the
semiconductor system
\begin{eqnarray}
\label{eq5}
&&\!\!\!\!\!\!\!\!\!\!\!\!
\partial_t n_{h,e} - \nabla \cdot {\left[ \pm {\mu _{h,e}
n_{h,e} \nabla (\phi - \phi _{N} ) + D_{h,e} \nabla n_{h,e}}  \right]} = 0,
\\ &&\!\!\!\!\!\!\!\!\!\!\!\!
\label{eq6}
\partial_t n_{\pm} - \nabla \cdot {\left[ \pm {\mu _{\pm}
n_{\pm} \nabla \phi  + D_{\pm} \nabla n_{\pm}}  \right]} = 0.
\end{eqnarray}
The labels denote holes ($h$), electrons ($e$),
 positive ($+$) and negative ($-$) ions. The electrostatic potential
$\phi $ is determined by Poisson's
equation $\varepsilon _{0} \nabla ^{2}\phi % = - {\sum\limits_{a} {q_{a} n_{a}}}
= -e(n_{ +}  + n_{h} - n_{ -}  - n_{e} )$. Still, in our system,
the quasi-neutrality condition $n_{ + } + n_{h} - n_{ -}  - n_{e} =
0$ is satisfied with very good accuracy. We note that the
quasi-neutrality condition does \textit{not} imply a constant
electric field strength, but rather indicates that small charge
imbalance causes extremely large electric fields \cite{16}.
%Formally, it may seem as if we now have an underdetermined system of
%equation. However, since the concentration of all species depends on
%$\phi $, the quasi-neutrality condition gives a constraint
%equation determining the potential \cite{16}.

%The treatment of the mobility with a general concentration dependence (see
%for instance Eq.\ (\ref{eq10}) below) will allow us to compare different models, and
%provide an important tool for understanding of the doping dynamics in organic semiconductors.
%As
%noted above, the introduction of the Nernst potential (2) in the doped
%region is justified due to the electrochemical nature of the system and the
%transformation of the polymer into a highly conducting metallic-like state
%when doped.
%The Nernst potential will contribute with a restoring force of
%the electromigration for the doping species due to the concentration
%dependence of the Fermi levels. We will show that a
%model without such an electrochemical potential contribution or a
%concentration dependent mobility cannot appropriately describe experimental observations (see Fig.\ 1).

In what follows we will focus on the dynamics and structure of the p-type
doping, but the analysis could of course equally well be applied to
the n-type doping. %From the above model we can derive some basic
%properties of the system.
Equations (\ref{eq5}) and (\ref{eq6}) admit a solution in a form of a localized planar doping front resembling a shock, moving with velocity $U$ and converting the undoped polymer into a doped polymer
(for a similar observation in ferroelectric crystals, see Ref. \cite{17}).
In the undoped region (label 0), we have a
uniform electric field $-\partial_x \phi _{0} $, uniform initial concentration of ions
$n_{ +0}  = n_{ -0}  \equiv n_{0} $, and no holes, $n_{h0} = 0$. Behind the
front, in the highly p-doped
region (label $\infty )$, the hole concentration is denoted by $
n_{h\infty},  $ %is determined by properties of the semiconductor, 
while the  electric field $-\partial_x \phi _{\infty}  $ and
ion concentrations $n_{\pm \infty}  $ are obtained from Eq.\ (\ref{eq5}) and the
quasi-neutrality condition. In the reference frame of the doping front
($\partial_t n = 0)$, we may integrate Eqs. (\ref{eq5}) and (\ref{eq6}) to obtain
\begin{eqnarray}
\label{eq7}
&&\!\!\!\!
D_{h} \partial_xn_{h} = - n_{h}U - n_{h} \mu _{h}
\partial_x(\phi - \phi _{N} ),
\\ &&\!\!\!\!
\label{eq8}
D_{\pm}  \partial_xn_{\pm} = (n_{0} - n_{\pm}  )U \pm \mu _{\pm}
\left( n_{\pm}  \partial_x\phi - n_{0} \partial_x\phi _{0} \right)  ,
\end{eqnarray}
for the holes and ions, respectively.
Noting that the concentration
gradients vanish on each side of the doping front, we obtain the p-front velocity
\begin{equation}
\label{eq9}
U = - (n_{0}/n_{h\infty})(\mu _{ +}  + \mu _{ -}
)\partial_x\phi _{0}.
\end{equation}
The electric field in the p-doped
region is negligible due to the high mobility of the holes
relative the ions. Diffusion does not influence the
front velocity (\ref{eq9}), but instead determines the structure and width
of the front. 

 A  dimensional
analysis suggests the characteristic
width of the front $L_{f}=D_{-}/U$. However,
$L_{f} $ does not portray the full structure of the
doping front, since there are several different characteristic length scales
within the front, from the undoped to the doped region.
The hole mobility is highly sensitive to
the concentration, and in the undoped region $\mu _{h0} /
\mu _{\pm}  \ll 1$. Thus, from Eqs.\ (\ref{eq7}) and (\ref{eq8}) we also find that the
characteristic length scale in the undoped region is given by $(\mu
_{h0} / \mu _{\pm}) L_{f}$, which may be orders of magnitude smaller than $L_f$.
As a result, one can expect a
sharp head of the doping front with strong concentration gradients. On the
other hand, for the doped region, Eqs.\ (\ref{eq7}), (\ref{eq8}) predict a slow variation of
the front due to $\mu _{h\infty}  / \mu
_{\pm}  \gg 1$ and the Nernst potential. The concentration approaches the final value asymptotically
according to the power law $(\mu _{ +}  / \mu _{ -}  + 1)L_{f} / (x-x_{f})$, where
$x-x_{f}$ measures the distance from the leading edge of the
doping front. This gives a very smooth behavior of the doping front in the
highly conducting part of the polymer. The analytical reasoning has been supported by the numerical solution to Eqs.\ (\ref{eq7}), (\ref{eq8}), described below and presented in
 Figs.\ 2, 3. Thus, the structure of the doping front and the generic features of the doping process are reproduced by
our model, both analytically and numerically. Some of the length scales presented above could be measured in future high resolution experiments.
%Thus, the predictions of
%our current model could shed further light on the doping process as
%measurement techniques are improved.

%%%% FIG 2 %%%%
\begin{figure}
	\includegraphics[width=0.98\columnwidth]{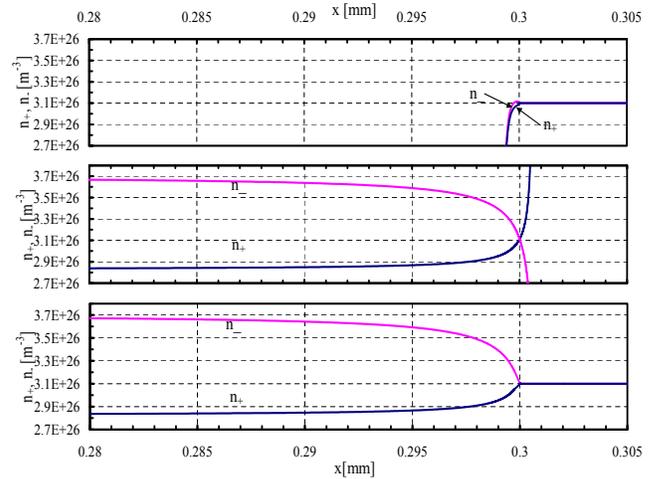}
	\caption{ Model data for the concentration of ions as a function of distance
from the positive anode, resulting from the employment of: (a)
concentration-\textit{dependent} mobility for the electronic charge carriers, (b)
concentration-\textit{independent} mobility and the Nernst equation in the doped regions, and
(c) concentration-\textit{dependent} mobility and the Nernst equation in the doped region.}
\end{figure}
%%%%%%%%%%%

%%%% FIG 3 %%%%
\begin{figure}
	\includegraphics[width=0.98\columnwidth]{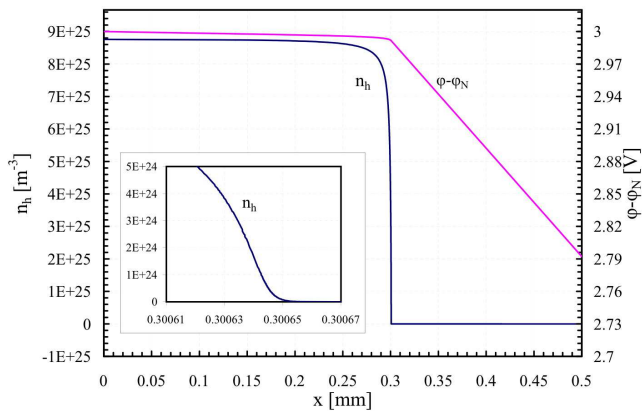}
	\caption{Model data showing concentration of holes, and the
overpotential profile as a function of distance from positive electrode.}
\end{figure}
%%%%%%%%%%%

As stated above, our model makes use of the concentration dependence of the
hole mobility (see, \textit{e.g.}, Refs.\ \cite{7,8}). 
In the numerical solution we employ the following empirical concentration-dependent mobility for holes as extracted from Ref.\ \cite{8}
%In the numerical solution we employ the
%empirical concentration-dependent mobility for  holes  extracted from
%Refs.\ \cite{8,18},
\begin{equation}
\label{eq10}
\mu _{h} = 3.85\times 10^{ - 8}{\left[ {1 + \tanh \left( {26.6\frac{n_{h}}{n_{0}} - 4.3} \right)} \right]}\, {\rm m}^{{\rm 2}} /{\rm V}{\rm s}{\rm }.
\end{equation}
Note that we have renormalized the mobility data in accordance with data from Ref.\ \cite{18} to account for the lower mobility of MEH-PPV in comparison to the conjugated polymer under study in Ref.\ \cite{19}.
To first approximation the ion mobility does not depend on concentration. To  analyze Eqs.\ (\ref{eq7}) and (\ref{eq8}) numerically, we
use Eq.\ (\ref{eq10}) and empirical data \cite{11,12,19} at the early stage of the doping
process: $n_{0} = 3.1 \times 10^{26}\,\mathrm{ m^{-3}}$, $d\phi_{0} / dx = 3
\,\mathrm{V/mm}$ , $\mu_{+} = 1.0\times 10^{-10}\, \mathrm{m^{2}/Vs}$, $\mu_{-} = 
2.2\times 10^{-10}\,\mathrm{ m^{2}/Vs}$, $n_{h\infty} = 8.8\times10^{25}
\, \mathrm{m^{-3}}$, and $T = 360\, \mathrm{K}$. The respective numerical solution is shown in Figs.\ 2 and 3. Figure 2
shows the ionic concentrations as a function of distance from the
anode ($x$) when the p-type doping front has progressed to $x = 0.3
\,\mathrm{mm}$. We note the highly resolved $x$-axis.
%Difference in ionic concentrations gives the
%doping concentration.
%, \textit{e.g.}, a specific p-type doping is achieved
%according to $n_{h} = n_{ -}  - n_{ +}  $.
Our model solution is shown in Fig.\ 2c, while Figs.\ 2a and b
illustrate the unphysical results without
employing the Nernst equation and concentration-dependent
mobility, respectively.
%Figures 2a and b demonstrate the fundamental importance
%of a combination of these two effects in order to obtain a physically sound
%solution.
First, by omitting the Nernst term, we obtain correct
concentrations in the undoped region but cannot achieve physically sound
results in the doped region (Fig.\ 2a). Second, if we
instead keep the Nernst term, and assume a constant
hole-mobility, the doped side of the front is
described quite well. However, ionic concentration levels diverge rapidly 
in the undoped
region (Fig.
2b).
%These two scenarios are clearly unphysical.
If we, on the other hand, include both the correct concentration dependent
mobility (\ref{eq10}) as well as the Nernst term (\ref{eq2}), we obtain the physically sound result shown in
Fig.\ 2c. Here we can see the emergence of a doping front that has a shape
in good agreement with the experimental observations of Fig.\ 1. The doping
front in Fig.\ 2c has a characteristic width $ \approx 0.1 - 0.2\,{\rm mm}$,
a consequence of the long tail due to the Nernst term in Eqs.
(\ref{eq8}) and (\ref{eq9}). At the leading edge of the front, the
gradients look very sharp as compared to the trailing edge. This
follows from the low hole mobility in the undoped region, and a
higher resolution of the leading edge demonstrates that the characteristic length
scale is $(\mu _{h0} / \mu _{\pm})L_{f}   \approx 2\times 10^{ - 4}{\rm
m}{\rm m}$, see the insert of Fig.\ 3.
Figure 3 shows the concentration of holes within the front, and the
overpotential profile as a function of distance from the anode. Again, in
agreement with the previous discussion, we observe a very sharp leading edge
of the front and a smooth long tail. The plot
demonstrates a weakly varying over-potential at the trailing edge of the front.
We also compare quantitatively the theoretical predictions for the
doping front velocity and the experimental measurements. Using
Eq.\ (\ref{eq9}) and the experimental data we calculate the front velocity at the
beginning of the doping process as $U = 13.5 \times 10^{ - 6}\,\mathrm{m / s}$.
The experimental measurements provided the same initial value for
the front velocity, $U = 13.5 \times 10^{ - 6}\,\mathrm{m / s}$, within the
accuracy of measurements.  As the doping progresses, the p- and
n-type doping fronts accelerate toward each other
\cite{15}, due to the decrease in distance between the
fronts.

Thus, our model is in very good agreement
with the experiments, both qualitatively and quantitatively.
It predicts the quantitative value of the doping
front velocity with good accuracy. The model also gives good agreement between the observed and calculated front thickness.
Furthermore, it reproduces the correct qualitative front structure as a
 combination of smooth tail and very sharp leading edge. Though the present analysis is one-dimensional, it can be easily extended to a multi-dimensional case.
However, the study of multi-dimensional effects is left for future research.

\acknowledgements

This work was supported by the Swedish Research Council under the under
Contracts No. 2007-4422, 2008-4422 and by the Kempe Foundation. LE is a Royal Academy of Sciences Research Fellow supported by a grant from the Knut and Alice Wallenberg Foundation.


\begin{thebibliography}{99}

\bibitem{1} A. J. Heeger, Rev. Mod. Phys. \textbf{73}, 681 (2001).

\bibitem{2} G. Malliaras, and R. Friend, Physics Today \textbf{58}, 53 (2005).

\bibitem{3} V. I. Arkhipov, P. Heremans, E. V. Emelianova, and H. Bassler, Phys. Rev. B \textbf{71} 045214 (2005).

\bibitem{4} V. I. Arkhipov, E. V. Emelianova, P. Heremans, and H. Bassler, Phys. Rev. B \textbf{72} 235202 (2005).

\bibitem{5} Q.\ Pei and O. Ingan\"as, Adv. Mater. \textbf{4}, 277 (1992).

\bibitem{6} J. H. Hou\textit{ et al.}, Chem. Commun., 6034 (2008).

\bibitem{7} X. Jiang\textit{ et al.}, Chem. Phys. Lett. \textbf{364},
616 (2002).

\bibitem{8} H. Shimotani, G. Diguet, and Y. Iwasa, Appl. Phys. Lett.
\textbf{86}, 022104 (2005).

\bibitem{9} L. Edman, Electrochimica Acta \textbf{50}, 3878 (2005).

\bibitem{Nature-Mat} P. Matyba, K. Maturova, M. Kemerink, N. Robinsion, L. Edman, Nature Materials \textbf{8}, 672 (2009).

\bibitem{10} J. Fang\textit{ et al.}, J. Am. Chem. Soc. \textbf{130}, 4562
(2008).

\bibitem{11} J. H. Shin\textit{ et al.}, Adv. Funct. Mater. \textbf{17}, 1807
(2007).

\bibitem{12} N. D. Robinson\textit{ et al.}, Phys. Rev. B \textbf{78}, 7
(2008).

\bibitem{13} J. F. Fang, Y. L. Yang, and L. Edman, Appl. Phys. Lett.
\textbf{93} 063503 (2008).

\bibitem{14} P. Matyba, M. R. Andersson, and L. Edman, Org. Electron.
\textbf{9}, 699 (2008).

\bibitem{15} N. D. Robinson, J. H. Shin, M. Berggren, L. Edman, Phys. Rev. B \textbf{74}, 155210
(2006).

\bibitem{16} F. Chen, \textit{Introduction to Plasma Physics}, Plenum Press,
NY, 1984.

\bibitem{17} S. Gronenborn, B. Sturman, M. Falk, D. Haertle, K. Buse, Phys.
Rev. Lett. \textbf{101}, 116601 (2008).

\bibitem{18} L.-L. Chua, J. Zaumseil, J.-F. Chang, E. Ou, P. Ho, H. Sirringhaus, R. H. Friend, Nature \textbf{434}, 194 (2005).

\bibitem{19} J.-H. Shin\textit{ et al.}, Appl. Phys. Lett. \textbf{89}, 013509 (2006).

\end{thebibliography}
\end{document}